\documentclass{article}
\usepackage{graphicx}

\begin{document}

\title{The Conduciveness of CA-rule Graphs}

\author{Valmir~C.~Barbosa\thanks{valmir@cos.ufrj.br.}\\
\\
Programa de Engenharia de Sistemas e Computa\c c\~ao, COPPE\\
Universidade Federal do Rio de Janeiro\\
Caixa Postal 68511\\
21941-972 Rio de Janeiro - RJ, Brazil}

\date{}

\maketitle

\begin{abstract}
Given two subsets $A$ and $B$ of nodes in a directed graph, the conduciveness of
the graph from $A$ to $B$ is the ratio representing how many of the edges
outgoing from nodes in $A$ are incoming to nodes in $B$. When the graph's nodes
stand for the possible solutions to certain problems of combinatorial
optimization, choosing its edges appropriately has been shown to lead to
conduciveness properties that provide useful insight into the performance of
algorithms to solve those problems. Here we study the conduciveness of CA-rule
graphs, that is, graphs whose node set is the set of all CA rules given a cell's
number of possible states and neighborhood size. We consider several different
edge sets interconnecting these nodes, both deterministic and random ones, and
derive analytical expressions for the resulting graph's conduciveness toward
rules having a fixed number of non-quiescent entries. We demonstrate that one of
the random edge sets, characterized by allowing nodes to be sparsely
interconnected across any Hamming distance between the corresponding rules, has
the potential of providing reasonable conduciveness toward the desired rules. We
conjecture that this may lie at the bottom of the best strategies known to date
for discovering complex rules to solve specific problems, all of an evolutionary
nature.

\end{abstract}

\newpage
\section{Introduction}

Ever since Wolfram first introduced his four-class qualitative categorization
of elementary cellular automata (CA) \cite{w84}, the problem of distinguishing
CA update rules in quantitative terms within both his classification scheme and
others (e.g., \cite{lp90,lpl90}), with the special aim of identifying the
so-called complex rules, has been a central one \cite{wl92,i01,w02,s09}. Some
of the notable approaches have been Langton's edge-of-chaos parameterization of
the rule space (through the fraction, denoted by $\lambda$, of ``non-quiescent''
entries in a rule) \cite{l90,lpl90} and Wuensche's input entropy (through
estimates, along traces of CA evolution, of the rate at which the various rule
entries are used) \cite{w99,bma06}. Despite criticism (e.g., \cite{mch94}),
these two approaches have remained emblematic because they have brought
important insight into the problem while occupying fundamentally different
niches: while the former attempts quantification by focusing on static
properties of the rule in question, the latter focuses on the rule's dynamic
response over time.

The larger issue, of course, is the identification of complex rules that display
specific patterns of behavior or solve specific problems, and in this regard all
classification-related quantifications seem to have had little impact. At
bottom, what really is behind the search for specific complex rules is an
intricate problem of combinatorial optimization that can easily become
unmanageable as the cells' possible states go beyond the binary case or their
neighborhoods get larger (either with the addition of extra dimensions or
otherwise). Not surprisingly, then, so far the success cases have all harnessed
nature-inspired stochastic methods, particularly those of evolutionary
inspiration \cite{mhc93,cmd03,rh05,bp10}, to navigate the rule space.

The use of ``navigate'' here is very appropriate because it evokes with great
clarity what combinatorial-optimization methods do, which is precisely to move
in a seemingly unstructured solution space seeking its optima. There is
structure, however, at least insofar as the method's optimization strategy can
be said to establish a relationship among the possible solutions as it moves
from one to another. There is also a more elemental type of structure connecting
the various solutions together, generally related to transforming one solution
into another by means of some simple alteration. Although this latter structure
need not be related to any given algorithm's navigation of the solution space,
for some problems it has been shown to provide the solution space with certain
``conduciveness'' characteristics that do nevertheless affect that algorithm's
performance \cite{b10}.

The problems in question are those of coloring an undirected graph's nodes
optimally and of finding one of the graph's largest subsets of nodes that only
contain non-neighbors (a so-called maximum independent set), both
computationally difficult in the sense of NP-hardness. For these two problems,
an underlying structure unrelated to the best existing heuristics has been shown
to account for intriguing performance transitions that are known to occur as the
graph's size changes. Specifically, right before such a transition it is
significantly harder to solve the problem than right past it. What happens at
the transition is that the aforementioned underlying structure suddenly becomes
much more conducive from nonoptimal to optimal solutions.

The notion of conduciveness we refer to is precise and can be formalized as
follows \cite{b10}. Let $D$ be a directed graph whose nodes stand for solutions
to the optimization problem at hand and whose edges reflect the said underlying
structure. Given two node subsets, call them $A$ and $B$, the conduciveness of
$D$ from $A$ to $B$ is the fraction of edges that, out of all those that are
outgoing from a node of $A$, are incoming to a node of $B$. Put differently, if
$m(A)$ is the number of edges whose tail nodes are in $A$, and $m(A,B)$ is the
number of edges with tail nodes in $A$ and head nodes in $B$, then the
conduciveness of $D$ from $A$ to $B$ is $m(A,B)/m(A)$. Conduciveness, then, is
necessarily a number in the $[0,1]$ interval, since every edge counted in
$m(A,B)$ is also counted in $m(A)$. In the two examples mentioned above, $A$ and
$B$ partition the node set of $D$ and stand, respectively, for nonoptimal and
optimal solutions to the optimization problem being considered.  

Here we examine the rule space of CA from the standpoint of some directed graphs
that can be viewed as providing an underlying structure interconnecting all
possible rules. As in the case of the graph problems mentioned above, such
structures need not have anything to do with possible algorithms to find
specific rules. Instead, we study their conduciveness properties in search for
some hint as to why evolutionary approaches to discover specific complex rules
have succeeded while others have barely been attempted. Our conclusions will
point at certain random structures whose expected conduciveness foreshadows the
existence of deterministic structures with the potential of being at least
reasonably conducive.

Of course, analyzing any graph's conduciveness requires a precise definition of
sets $A$ and $B$. In the case of CA this can be really tricky. Say, for
example, that we are looking for a complex rule to solve a specific problem.
Sets $A$ and $B$ might then be defined as a function of some quantitative
description of how well each possible rule solves that problem. This would
amount to simply carrying over, to the context of CA rule space, the very same
simulation-based approach that was used in the graph-coloring and
independent-set problems mentioned above. While we had success in those cases,
mainly because scaled-down versions of the problems still exhibit the same
transition phenomena we wished to explain, nothing of the sort is expected to
happen in the case of CA. In other words, we would be left with impossibly large
rule spaces and would never be able to characterize conduciveness properly.

The alternative we adopt in this paper is to settle for some characterization of
the rule space which, while retaining the ability to relate to a rule's
``complexity'' to some extent, is also amenable to an analytical portrayal of
conduciveness that can be used in lieu of computer simulations. The advantage,
clearly, is that entire rule spaces can be examined, at least in some nontrivial
cases. Our choice has been to use Langton's $\lambda$ parameter, so rules in set
$B$ are characterized by having the same number of non-quiescent entries. Set
$A$ is then the complement of $B$ with respect to the entire rule space. The
disadvantage we have to cope with is, naturally, the loss in power to describe
complexity that has been an issue also in Langton's approach.

We proceed in the following manner. First we introduce, in
Section~\ref{sec:nets}, the CA-rule graphs to be studied. Then we derive
analytical expressions for their conduciveness in Section~\ref{sec:cond} and
study them with the aid of selected plots in Section~\ref{sec:plots}. We discuss
the most relevant properties and finds in Section~\ref{sec:disc} and conclude in
Section~\ref{sec:concl}.

\section{CA-rule graphs}\label{sec:nets}

We consider CA in which a cell's state is one of the integers in
$\{0,1,\ldots,s-1\}$ for some $s\ge 2$. We assume that the cell's neighborhood,
including the cell itself, has size $\delta$ for some $\delta\ge 2$. It follows
that the rule governing the behavior of the CA can be regarded as an $L$-entry
table for $L=s^\delta$ and that the number of possible rules is $s^L$. Cells may
be arranged with respect to one another one-dimensionally or otherwise, as this
is of no concern to what follows.

We focus on the directed graph having one node for each possible rule and edges
that join nodes according to one of three criteria. Two of them are
deterministic and result in an edge existing from one node to another if and
only if that edge's antiparallel counterpart also exists. Using an undirected
graph instead would then be entirely acceptable, but we refrain from doing so to
adhere to the definition of conduciveness and to maintain compatibility with the
third, probabilistic criterion.

The first criterion joins two nodes if and only if the corresponding rules
differ in exactly one entry (i.e., if the Hamming distance between them is
exactly $1$). This is the case of the traditional hypercube, which we denote by
$H$. In $H$ every node has exactly $L(s-1)$ out-neighbors. The second criterion
generalizes the first one by allowing two nodes to be joined if and only if the
Hamming distance between the corresponding rules is exactly $h$ for some
$h\ge 1$. The resulting graph is a generalized hypercube, here denoted by $H^+$.
In $H^+$ every node has ${L\choose h}(s-1)^h$ out-neighbors, since this is the
number of ways in which its rule can be modified by altering exactly $h$
entries.

The third criterion to define the graph's edge set is to allow any two nodes to
be joined probabilistically to each other as a function of the Hamming distance
between their rules. This is done independently for each of the two possible
directions, so two nodes need no longer be joined by an antiparallel edge pair.
The result is a random-graph model of the interconnections among the rules. This
random graph is denoted by $H^\mathrm{r}$ and depends on a probability
parameter, call it $p$. In $H^\mathrm{r}$ an edge exists from a node to another
with probability $p^h$, where $h$ is the Hamming distance between the nodes'
rules. That is, although any Hamming distance is allowed between the rules of
two nodes joined by an edge, higher Hamming distances make it exponentially less
likely that the edge indeed exists. For fixed $h\ge 1$ we expect a node to have
$p^h{L\choose h}(s-1)^h$ out-neighbors separated from it by a Hamming distance
of $h$, so overall the expected number of a node's out-neighbors is
\begin{equation}
\sum_{h=1}^L{L\choose h}[p(s-1)]^h=\left[p(s-1)+1\right]^L-1.
\end{equation}

For each of $H$, $H^+$, and $H^\mathrm{r}$, and for each $\ell$ such that
$0\le\ell\le L$, we partition the graph's node set into the two sets $A$ and
$B$, the latter containing all (and only) nodes whose rules have exactly $\ell$
non-quiescent entries. It follows that $B$ comprises ${L\choose\ell}(s-1)^\ell$
nodes. We then calculate each graph's conduciveness from set $A$ to set $B$,
denoted respectively by $C_\ell$, $C_\ell^+$, and $C_\ell^\mathrm{r}$. Owing to
the random nature of $H^\mathrm{r}$, $C_\ell^\mathrm{r}$ is the expected
conduciveness from $A$ to $B$.

\section{Conduciveness formulae}\label{sec:cond}

We begin with the hypercube $H$. In this case the total number of edges outgoing
from nodes in set $A$ is the product of the set's cardinality and the number of
out-neighbors of each of its nodes, that is,
$[s^L-{L\choose\ell}(s-1)^\ell]L(s-1)$. Some of these edges are incoming to
nodes in set $B$, belonging to one of two categories.

Edges in the first category outgo from nodes of $A$ whose rules have exactly one
non-quiescent entry too few when compared to those of $B$, provided $\ell>0$.
The number of such nodes is ${L\choose{\ell-1}}(s-1)^{\ell-1}$, each one
accounting for $(L-\ell+1)(s-1)$ $B$-bound edges, since $s-1$ is the number of
possibilities to turn each of the $L-\ell+1$ quiescent entries into a
non-quiescent one. The second category of $B$-bound edges comprises edges
outgoing from nodes in $A$ that have exactly one non-quiescent entry too many
with respect to $B$, provided $\ell<L$. There are
${L\choose{\ell+1}}(s-1)^{\ell+1}$ such nodes, each one contributing $\ell+1$ to
the total of $B$-bound edges, this being the number of non-quiescent entries,
each affording one single possibility to be turned into a quiescent one. It then
follows that $C_\ell$ is given by
\begin{equation}
C_\ell=
\frac
{{\displaystyle{\delta_{\ell>0}{L\choose{\ell-1}}(s-1)^{\ell-1}(L-\ell+1)(s-1)+
\mbox{}\hspace{0.5in}}\atop
{\hfill\displaystyle\delta_{\ell<L}{L\choose{\ell+1}}(s-1)^{\ell+1}(\ell+1)}}}
{\displaystyle\left[s^L-{L\choose\ell}(s-1)^\ell\right]L(s-1)},
\label{eq:cond}
\end{equation}
where each of $\delta_{\ell>0}$ and $\delta_{\ell<L}$ equals $1$ if the
corresponding inequality holds, $0$ otherwise.

As we move to the generalized hypercube $H^+$, the number of edges outgoing from
nodes in $A$ becomes $[s^L-{L\choose\ell}(s-1)^\ell]{L\choose h}(s-1)^h$, and we
are left with the task of calculating how many of them are incoming to nodes in
$B$. Again we categorize these edges as a function of their end nodes on the $A$
side, but now we require a nonnegative integer parameter, call it $k$, to
proceed.

Each value of $k$ corresponds to nodes in $A$ whose rules have exactly
$\ell-h+2k$ non-quiescent entries, and consequently $L-\ell+h-2k$ quiescent
entries (provided $k\neq h/2$, in which case we would have a $B$ node, not an
$A$ node). Simultaneously altering $h$ entries, $k$ of them from non-quiescent
to quiescent and the remaining $h-k$ from quiescent to non-quiescent, clearly
leads to a node in $B$, since the number of non-quiescent entries is thus
changed to $\ell$ by the subtraction of $k-(h-k)$ off the original value,
$\ell-h+2k$. We denote the number of such nodes in $A$ by $f(k)$, therefore
\begin{equation}
f(k)={L\choose{\ell-h+2k}}(s-1)^{\ell-h+2k}.
\end{equation}
Each of these nodes allows for
${{L-\ell+h-2k}\choose{h-k}}{{\ell-h+2k}\choose k}$ possibilities to effect the
said alterations, each possibility accounting for $(s-1)^{h-k}$ $B$-bound edges.
Denoting by $g(k)$ the overall number of $B$-bound edges outgoing from a given
node in $A$ yields
\begin{equation}
g(k)={{L-\ell+h-2k}\choose{h-k}}{{\ell-h+2k}\choose k}(s-1)^{h-k}.
\end{equation}

We then have
\begin{equation}
C_\ell^+=
\frac
{\displaystyle
\sum_{{\scriptstyle k=\max\{0,h-\ell\}}\atop{\scriptstyle k\neq h/2}}
^{\min\{h,L-\ell\}}
f(k)g(k)}
{\displaystyle\left[s^L-{L\choose\ell}(s-1)^\ell\right]{L\choose h}(s-1)^h},
\label{eq:cond+}
\end{equation}
where the possible values of $k$ are carefully controlled to account for the
forbidden cases of $k\notin[0,h]$ and $k=h/2$. Note, incidentally, that letting
$h=1$ causes the numerator of Eq.~(\ref{eq:cond+}) to have at most two summands,
one for $k=0$ and one for $k=1$, in such a way that $f(0)g(0)$ and $f(1)g(1)$
are precisely the summands in the numerator of Eq.~(\ref{eq:cond}),
respectively the leftmost one and the rightmost.

In the case of the random graph $H^\mathrm{r}$, the expected number of edges
outgoing from nodes in set $A$ is
$[s^L-{L\choose\ell}(s-1)^\ell]\{[p(s-1)+1]^L-1\}$. We calculate how many of
these edges are expected to be $B$-bound by simply summing up, on $h$, the
corresponding number we found in the case of the generalized hypercube $H^+$
(i.e., for the fixed Hamming distance $h$). In this sum every edge is weighted
by the probability $p^h$ that defines its existence. We obtain
\begin{equation}
C_\ell^\mathrm{r}=
\frac
{\displaystyle\sum_{h=1}^Lp^h
\sum_{{\scriptstyle k=\max\{0,h-\ell\}}\atop{\scriptstyle k\neq h/2}}
^{\min\{h,L-\ell\}}
f(k)g(k)}
{\displaystyle\left[s^L-{L\choose\ell}(s-1)^\ell\right]
\left\{\left[p(s-1)+1\right]^L-1\right\}}.
\label{eq:condr}
\end{equation}

Note that, in the limit as $p\to 0$, $C_\ell^\mathrm{r}$ tends to $C_\ell^+$ for
$h=1$, that is, the conduciveness $C_\ell$ of the hypercube $H$. To see this,
first notice that, as the limit is approached, the only value of $h$ still
contributing to the numerator of Eq.~(\ref{eq:condr}) is $h=1$. The resulting
simplification leads to Eq.~(\ref{eq:cond}) through Eq.~(\ref{eq:cond+}), once
we realize that
\begin{equation}
\lim_{p\to 0}\frac{[p(s-1)+1]^L-1}{p}=L(s-1).
\label{eq:lim}
\end{equation}

\section{Conduciveness plots}\label{sec:plots}

In this section we present plots of the hypercube conduciveness $C_\ell$, the
conduciveness $C_\ell^+$ of the generalized hypercube, and the random-graph
conduciveness $C_\ell^\mathrm{r}$, as per Eqs.~(\ref{eq:cond}),
(\ref{eq:cond+}), and~(\ref{eq:condr}), respectively. In all plots we normalize
the abscissae to lie in the $[0,1]$ interval by plotting the conduciveness
values against $\lambda=\ell/L$, the Langton parameter.

Conduciveness values can be extremely low, depending on the parameters involved,
which requires some care in both handling the generation of the data to be
plotted and the plotting itself, and even so constrains the parameter values
that can be used. We have used a C program to generate the data as
\texttt{long double} numbers (96-bit numbers for \texttt{gcc-4.4.6-3}) and
\texttt{gnuplot-4.2.6-2} to do the actual plotting. As \texttt{gnuplot-4.2.6-2}
does not appear to handle numbers of the same precision as those we generated
via \texttt{gcc-4.4.6-3}, and also to avoid the use of an automatic logarithmic
scale while plotting (we think this facilitates reading figures off the plots),
a conduciveness value $c$ is output as $\mathrm{LL}(c)=\log_{10}(-\log_{10}c)$
for plotting. That is, reading an ordinate $\mathrm{LL}(c)=y$ off a plot implies
a conduciveness value $c=10^{-10^y}$.

\begin{figure}[p]
\centering
\scalebox{0.80}{\includegraphics{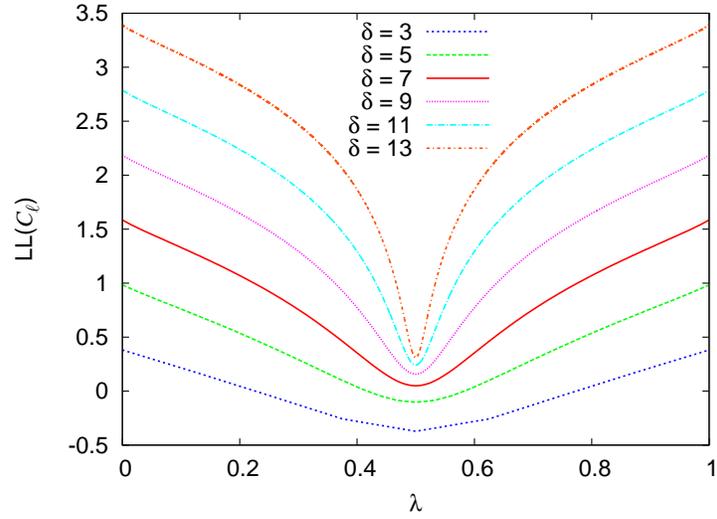}}
\caption{Conduciveness $C_\ell$ of the hypercube $H$ for $s=2$. Data are given
against $\lambda=\ell/L$.}
\label{fig1}
\end{figure}

\begin{figure}[p]
\centering
\scalebox{0.80}{\includegraphics{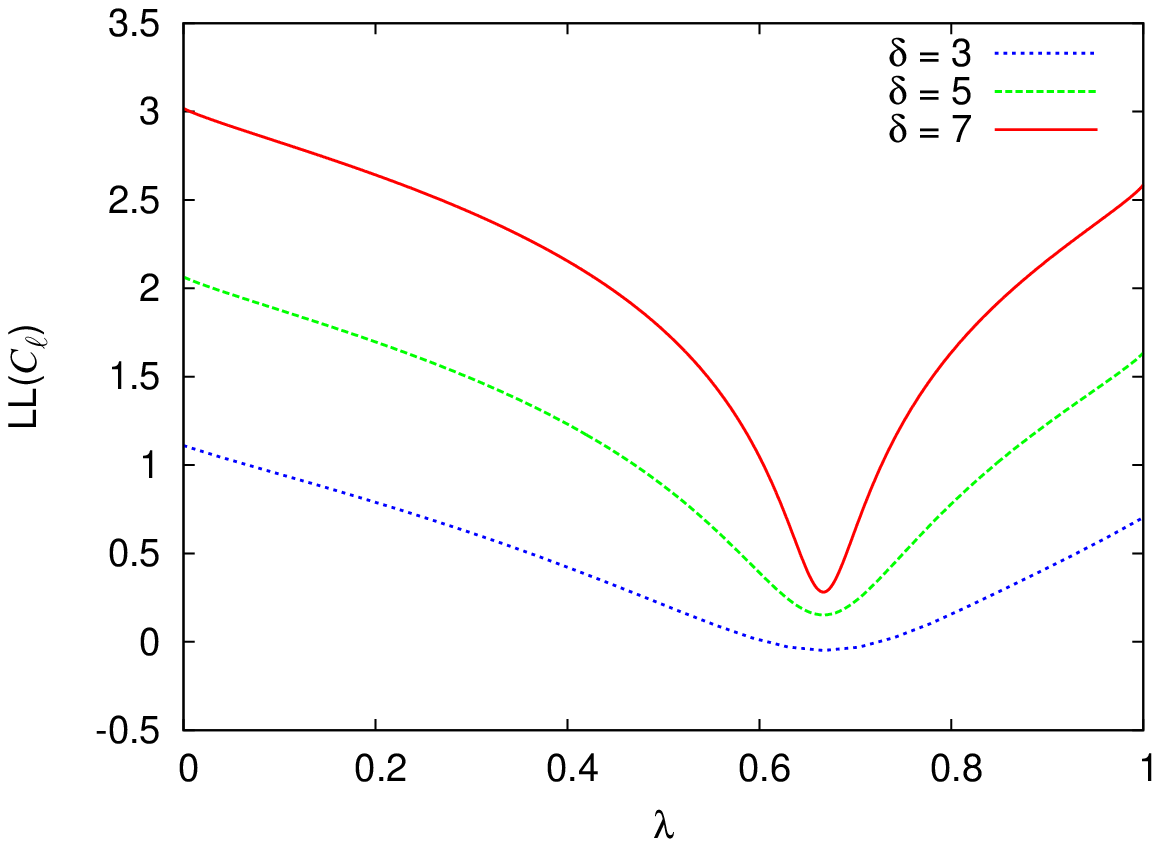}}
\caption{Conduciveness $C_\ell$ of the hypercube $H$ for $s=3$. Data are given
against $\lambda=\ell/L$.}
\label{fig2}
\end{figure}

\begin{figure}[p]
\centering
\scalebox{0.80}{\includegraphics{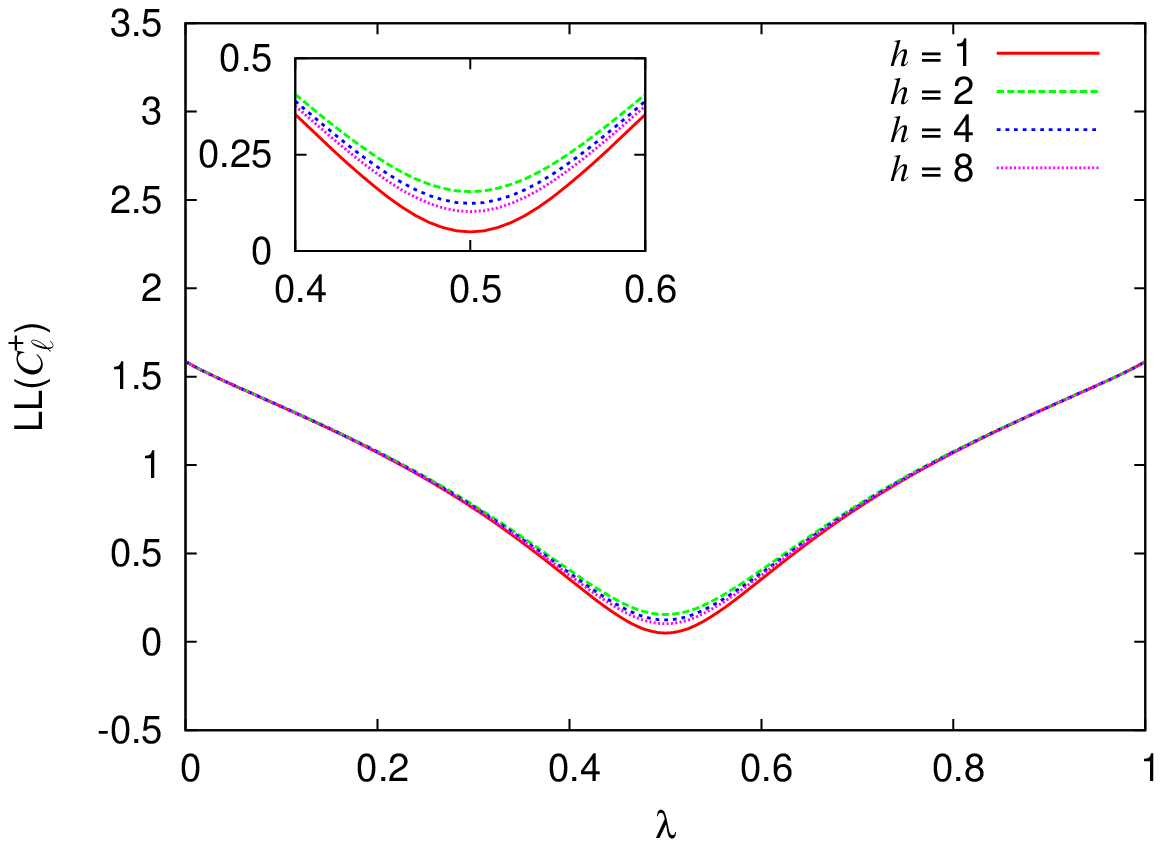}}
\caption{Conduciveness $C_\ell^+$ of the generalized hypercube $H^+$ for $s=2$
and $\delta=7$. Data are given against $\lambda=\ell/L$.}
\label{fig3}
\end{figure}

\begin{figure}[p]
\centering
\scalebox{0.80}{\includegraphics{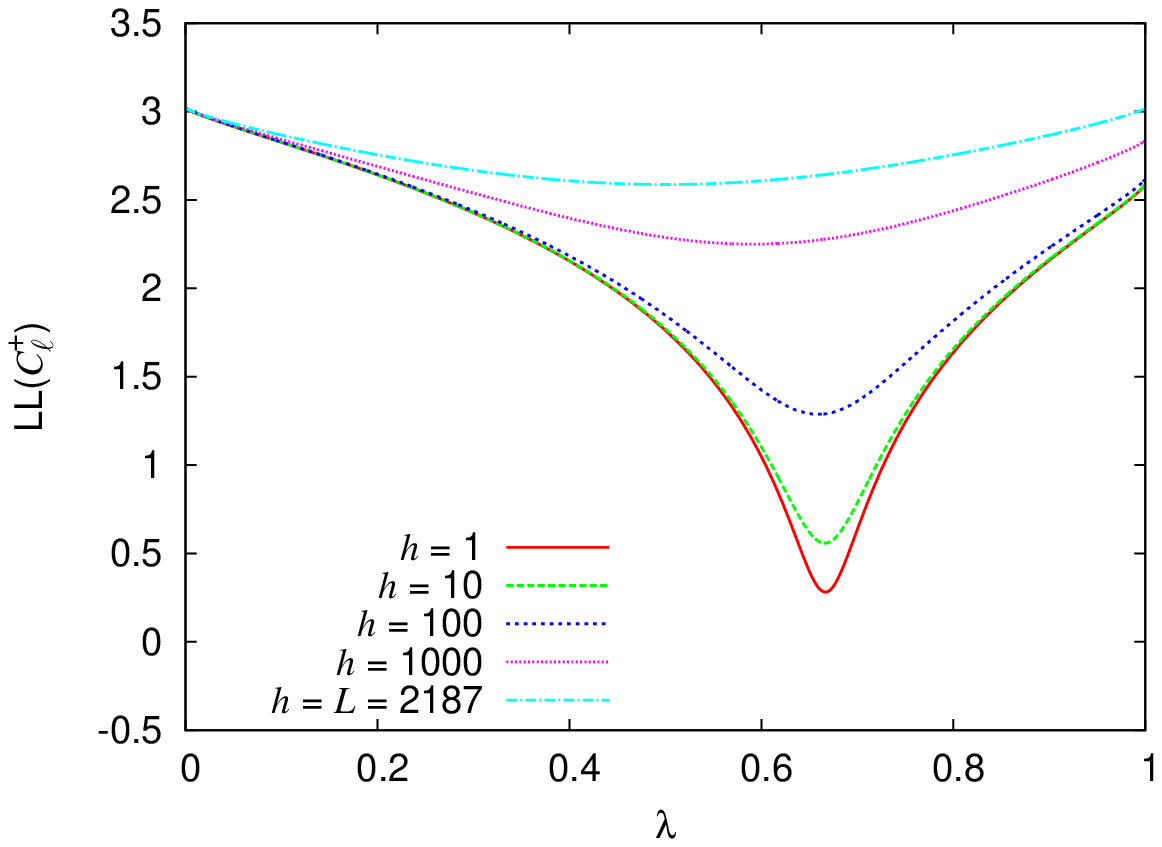}}
\caption{Conduciveness $C_\ell^+$ of the generalized hypercube $H^+$ for $s=3$
and $\delta=7$. Data are given against $\lambda=\ell/L$.}
\label{fig4}
\end{figure}

\begin{figure}[p]
\centering
\scalebox{0.80}{\includegraphics{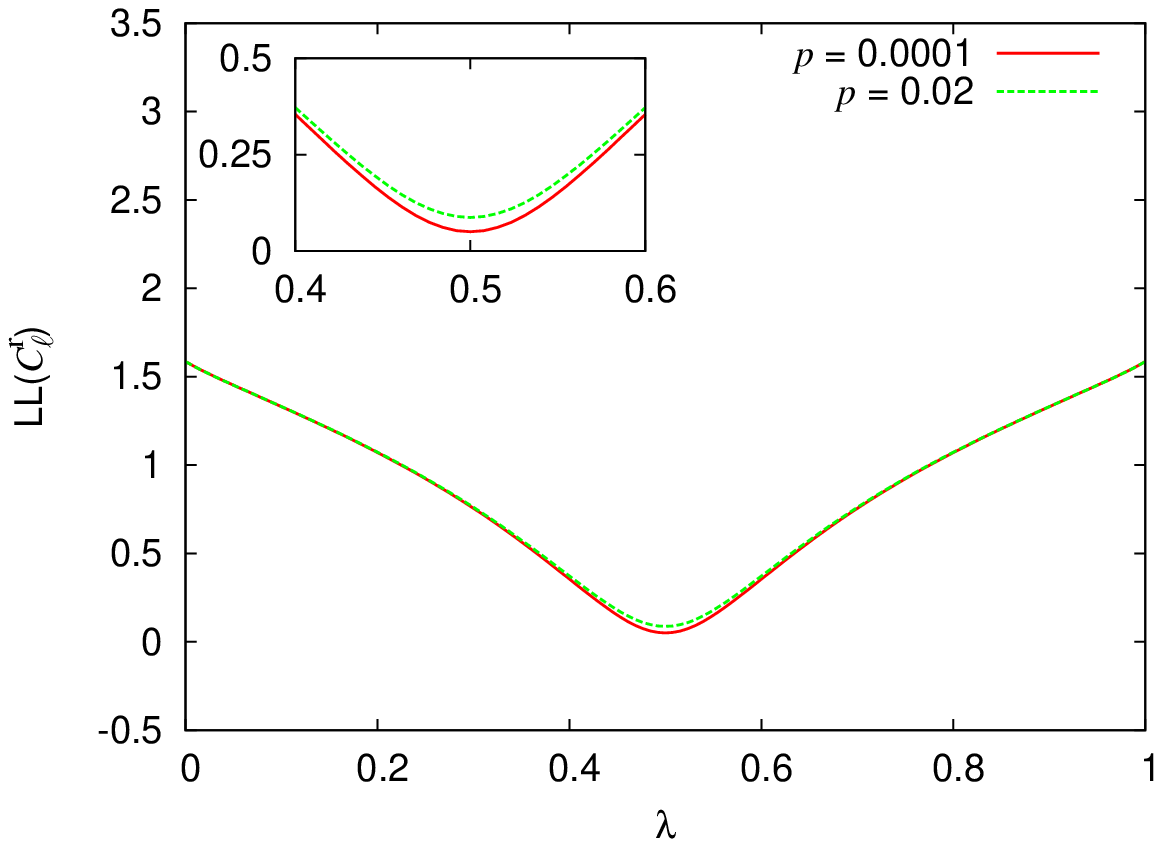}}
\caption{Expected conduciveness $C_\ell^\mathrm{r}$ of the random graph
$H^\mathrm{r}$ for $s=2$ and $\delta=7$. Data are given against
$\lambda=\ell/L$.}
\label{fig5}
\end{figure}

\begin{figure}[p]
\centering
\scalebox{0.80}{\includegraphics{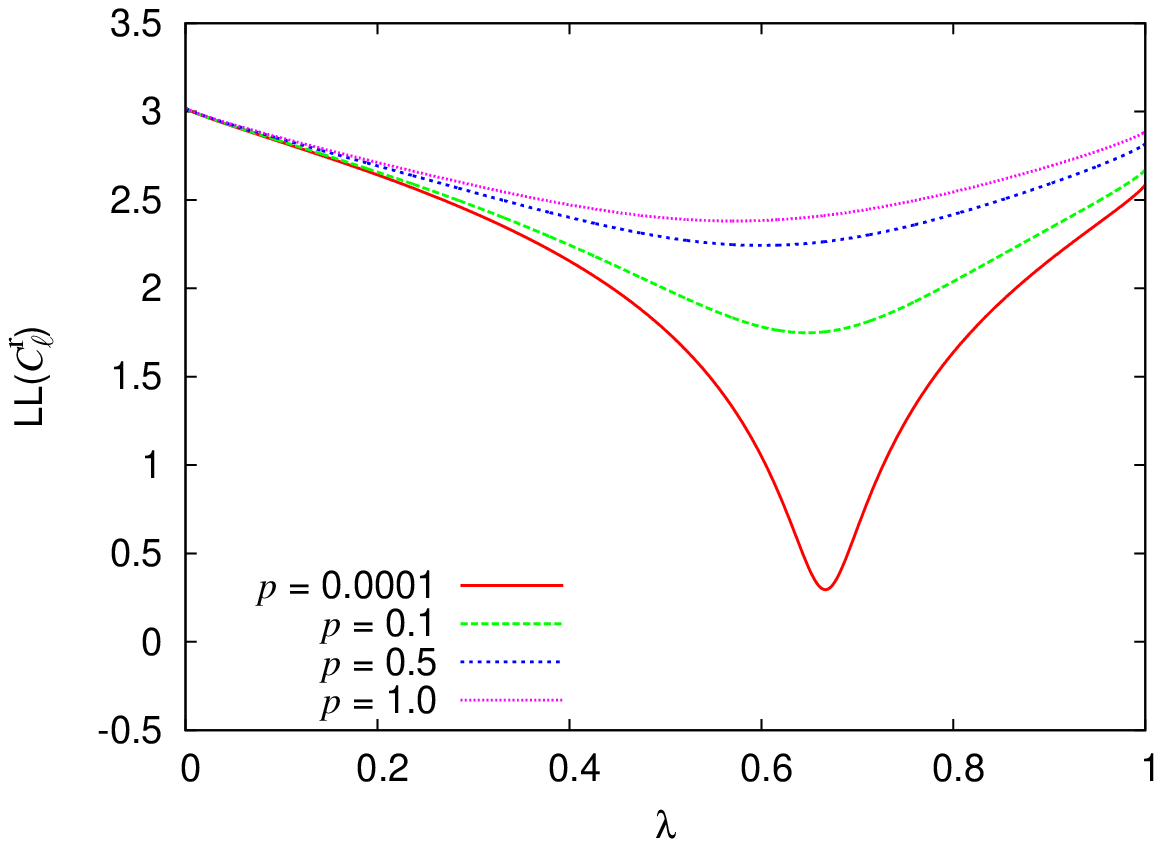}}
\caption{Expected conduciveness $C_\ell^\mathrm{r}$ of the random graph
$H^\mathrm{r}$ for $s=3$ and $\delta=7$. Data are given against
$\lambda=\ell/L$.}
\label{fig6}
\end{figure}

Plots for $C_\ell$ are shown in Figures~\ref{fig1} and~\ref{fig2} for $s=2$ and
$s=3$, respectively, and a variety of $\delta$ values. Plots for $C_\ell^+$ are
given in Figures~\ref{fig3} and~\ref{fig4}, respectively for $s=2$ and $s=3$ as
well, now for $\delta$ fixed at $\delta=7$ with a variety of $h$ values. Plots
for $C_\ell^\mathrm{r}$ appear in Figures~\ref{fig5} and~\ref{fig6}, once again
for $s=2$ and $s=3$, respectively, again for $\delta=7$ but now varying $p$. All
three figures corresponding to the same value of $s$ have one plot in common:
the $C_\ell$ plot for $\delta=7$, which is the same as the $C_\ell^+$ plot for
$\delta=7$ with $h=1$, which in turn is visually indistinguishable from the
$C_\ell^\mathrm{r}$ plot for $\delta=7$ with $p=0.0001$ (by virtue of the limit
given in Eq.~(\ref{eq:lim})). For ease of reference, note that the integer
ordinates $0$, $1$, $2$, and $3$ appearing in all figures correspond to
conduciveness values of $10^{-1}$, $10^{-10}$, $10^{-100}$, and $10^{-1000}$,
respectively.

\section{Discussion}\label{sec:disc}

One common term in all of Eqs.~(\ref{eq:cond}), (\ref{eq:cond+}), and
(\ref{eq:condr}) is the number of nodes whose rules contain exactly $\ell$
non-quiescent entries, given by ${L\choose\ell}(s-1)^\ell$. It is easy to prove
that this number is maximized by choosing $\ell=\ell^*$, where
\begin{equation}
\frac{\ell^*}{L}=\left(1-\frac{1}{s}\right),
\end{equation}
which is precisely the probability of picking a non-quiescent entry in a rule
where all $s$ values are equally represented. In his analysis of elementary CA
\cite{l90,wl92}, Langton associated the resulting $\lambda^*=\ell^*/L$ with the
occurrence of chaotic behavior. Moreover, deviating from the optimal value to
either side might lead to complex rules and eventually to trivial fixed points
and limit cycles.

As it happens, it can also be proven that setting $\ell=\ell^*$ maximizes
$C_\ell$ as well. This is illustrated clearly in Figures~\ref{fig1}
and~\ref{fig2}, where $\lambda^*=0.5$ in the former case and $\lambda^*=2/3$ in
the latter, regardless of the value of $\delta$. Thus, if Langton's scheme were
to hold as originally proposed, the hypercube $H$ would be much more conducive
to chaotic-rule nodes than to those of rules leading to fixed points or limit
cycles, with the conduciveness to complex-rule nodes lying somewhere in between.

Figures~\ref{fig1} and~\ref{fig2} also reveal that, for fixed $\delta$, the
value of $C_\ell$ falls quickly as $\ell$ is moved to either side of its optimal
value, $\ell^*$. In fact, this fall eventually leads to staggeringly low
conduciveness values for the higher values of $\delta$. Curiously, though, for
$\ell=\ell^*$ the decrease in $C_\ell$ for increasing $\delta$ seems headed
toward a limiting value. However, this can be seen to be illusory by examining
the case of $s=2$ (thus $\ell^*=L/2=2^{\delta-1}$). In this case, we can rewrite
$C_{\ell^*}$ as
\begin{equation}
C_{2^{\delta-1}}=
\frac{1}
{\frac{\displaystyle2^{2^\delta}}
{\displaystyle{2^\delta\choose 2^{\delta-1}}}-1},
\end{equation}
whose limit as $\delta\to\infty$ is infinity.

The generalized hypercube $H^+$, to which Figures~\ref{fig3} and~\ref{fig4}
refer, represents an attempt to increase a node's number of out-neighbors in the
graph from the $L(s-1)$ out-neighbors that it has in the hypercube $H$ to
${L\choose h}(s-1)^h$ for $h>1$. This increase is not steady with $h$, though:
as in the characterization of $\ell^*$ above, this number of out-neighbors peaks
at $h=L(1-1/s)$ and then decreases as $h$ continues to grow toward $h=L$.

In any event, Figures~\ref{fig3} and~\ref{fig4} indicate that $C_\ell^+$ does
not improve with respect to $C_\ell$ by simply increasing the Hamming distance
between the rules of two interconnected nodes. On the contrary, as $s$ is
increased from $2$ to $3$ we see that conduciveness values worsen dramatically
as $h$ is increased, in a clear indication that $h=1$ remains the best choice.
We also remark that, although for $s=3$ the lowering of $C_\ell^+$ values occurs
monotonically with the increasing of $h$, the case of $s=2$ is altogether
different. Specifically, all $C_\ell^+$ values are confined between those for
$h=1$ and $h=2$, with those for odd $h$ coinciding with those of $h=1$ and those
for even $h$ increasing steadily toward those of $h=1$ as well (this can be seen
more clearly in the inset to Figure~\ref{fig3}).

Similar observations apply to the random graph $H^\mathrm{r}$. Note initially
that here too there has been an attempt to increase a node's number of
out-neighbors in the graph, though in the sense of probabilistic expectation and
allowing a random mixture of Hamming distances between a node's rule and those
of its out-neighbors. In fact, this expected number of out-neighbors, given by
$[p(s-1)+1]^L-1$, can be seen to increase steadily with increasing $p$.
However, increasing the expected number of out-neighbors of a node does not
contribute to improve the behavior of $C_\ell^\mathrm{r}$, whose values are seen
to fall precipitously as $p$ is increased for $s=3$ (cf.\ Figure~\ref{fig6}).
The case of $s=2$, shown in Figure~\ref{fig5}, is sort of an oddity, with all
conduciveness values confined between those for a very low value of $p$ and
those for about $p=0.02$. We show no further plots than those of these
constraining values of $p$ to avoid cluttering the figure, but remark that
$C_\ell^\mathrm{r}$ first decreases as $p$ is increased from $p=0.0001$, then
increases back toward its initial value after $p=0.02$ is reached.

It might then seem like the best conduciveness is provided by graph $H$, the
hypercube, since $C_\ell\ge C_\ell^+$ for any value of $h$ and
$C_\ell\ge C_\ell^\mathrm{r}$ for any value of $p$. The caveat, of course, is
that the latter inequality requires careful interpretation, since
$C_\ell^\mathrm{r}$ is the expected conduciveness of all graphs modeled by the
random graph $H^\mathrm{r}$, not the conduciveness of a specific graph. The
graphs to which the expected value refers include any graph one may come up
with, because $H^\mathrm{r}$ allows edges to exist between any two nodes, in any
of the two possible directions, regardless of the Hamming distance between their
rules. This means that the conduciveness distribution to which the expected
value refers, although unknown, spreads toward lower conduciveness values very
widely, as shown in Figure~\ref{fig6} for $s=3$. The inescapable conclusion is
that $H^\mathrm{r}$ also models graphs whose conduciveness is higher than
$C_\ell$. All we know about these graphs, though, is that they allow mixed
Hamming distances between interconnected nodes' rules to coexist and that the
best improvements in conduciveness should occur for low values of $p$.

Allowing diverse Hamming distances to occur in the same graph is more of a key
property of $H^\mathrm{r}$ than it may at first seem. To see that this is so,
let us consider another random-graph model, viz.\ a directed variation of the
Erd\H{o}s-R\'{e}nyi model \cite{er59,k90}, henceforth referred to as DER. In
this model, an edge exists between any two distinct nodes, in each of the two
possible directions, independently with probability $p$. In our setting this
leads to an expected number of out-neighbors of $p(s^L-1)$. The expected
conduciveness of the DER model can be obtained from that of $H^\mathrm{r}$ in
Eq.~(\ref{eq:condr}) by substituting $p$ for $p^h$ in the numerator and
$p(s^L-1)$ for $[p(s-1)+1]^L-1$ in the denominator. The resulting expression is
independent of $p$, being in fact identical to $C_\ell^\mathrm{r}$ for $p=1$.
The latter, of course, is precisely the special case of $H^\mathrm{r}$ that is
no longer a random graph but the complete graph instead, that is, the graph in
which every node has every other node as an out-neighbor. So, although the DER
graph also allows for conduciveness values that spread around the expected value
and in fact encompass the conduciveness of any other graph, this expected value
is as bad as the conduciveness of $H^\mathrm{r}$ for $p=1$. Therefore the two
random-graph models, $H^\mathrm{r}$ for low values of $p$ and DER, have expected
conduciveness values corresponding to the upper and lower conduciveness extremes
of Figure~\ref{fig6}, respectively.

\section{Conclusions}\label{sec:concl}

Applying the notion of a graph's conduciveness when the graph's node set is the
solution space of some combinatorial problem and its edge set reflects some
elemental relationship among the various solutions is a technique for
discovering whether the graph possesses some inherent property that explains
the behavior of algorithms to search for specific nodes in it. The idea is very
new, dating from its first use in \cite{b10}, so it is no surprise that we have
little more than a phenomenological understanding of how conduciveness relates
to search algorithms that in general use totally different sets of edges while
seeking nodes belonging to a particular set, say $B$. One tantalizing
interpretation is that, as such an algorithm traverses the node set,
occasionally the two edge sets will coincide and, if the graph is conducive
toward $B$ from outside $B$, then the possibility of reaching $B$ presents
itself.

The study contained in \cite{b10} seems to support this interpretation, and so
does the present one, which has been about traversing the rule space of CA
searching for some degree of complexity which, for the sake of permitting an
analytical formulation of conduciveness in all graph types investigated, we
assumed to be related to the rules' density of non-quiescent entries. Our main
conclusion has been that a sparse random-graph topology allowing nodes to be
interconnected regardless of the Hamming distance separating the rules they
stand for has the potential of providing reasonable conduciveness toward the
desired rules, particularly if these rules' number of non-quiescent entries is
located not too far from $L(1-1/s)$ in the sequence $0,1,\ldots,L$. We think
this may be well in line with the success of some evolutionary approaches in
locating complex rules to solve specific problems: though the recombine/mutate
essence of such approaches leads them to follow routes of their own through rule
space, its stochastic character is bound to allow for successful jumps into set
$B$ whenever the expected conduciveness is sufficiently high.

We finalize by noting that conduciveness studies like this one also constitute
a link between the study of CA and that of the so-called complex networks, which
over the past decade have been applied so successfully to such a wide range of
domains as reported in \cite{bs03,nbw06,bkm09}. As demonstrated by the recent
study in \cite{k10}, the field of artificial life has much to gain from the
broadly applicable, essentially stochastic tools that researchers on complex
networks have amassed for the analysis of very large ensembles of interconnected
elements. Our study of the conduciveness of CA-rule graphs constitutes another
example.

\subsection*{Acknowledgments}

We acknowledge partial support from CNPq, CAPES, and a FAPERJ BBP grant.

\bibliography{condca}
\bibliographystyle{plain}

\end{document}